# Pokémon Go: Impact on Yelp Restaurant Reviews

Pavan Ravikanth Kondamudi
Northern Illinois University
DeKalb, USA
pkondamudi1@niu.edu

Bradley Protano
Northern Illinois University
DeKalb, USA
bprotano1@niu.edu

Hamed Alhoori
Northern Illinois University
DeKalb, USA
alhoori@niu.edu

## ABSTRACT
Pokémon Go, the popular Augmented Reality based mobile application, launched in July of 2016. The game's meteoric rise in usage since that time has had an impact on not just the mobile gaming industry, but also the physical activity of players, where they travel, where they spend their money, and possibly how they interact with other social media applications. In this paper, we studied the impact of Pokémon Go on Yelp reviews. We separated restaurants by those that had PokéStops near them and those that did not. For restaurants near PokéStops, we found a slight drop in the number of reviews.

## CCS CONCEPTS
• Human-centered computing → Collaborative and social computing → Empirical studies in collaborative and social computing.

## KEYWORDS
Pokemon Go; PokéStops; Augmented Reality Games; Yelp; Local Businesses; Restaurants; Reviews; Social Media.

## 1. INTRODUCTION
Augmented reality games allow users to play in a mediated reality produced by imposing the graphics from the game onto a real-world environment. This technology is having an impact on the world in social, cultural, and economic ways. Pokémon Go is one such popular augmented reality game that was built around the famous Pokémon, or "Pocket Monsters", series in which a player catches the titular creatures using PokéBalls. In this game, players move from one place to another with their gaming devices looking for these Pokémon. Players are provided with a limited number of PokéBalls initially, although when they run out, they can collect more of these and other resources from PokéStops. The PokéStops are usually located at places where people often gather in relatively large numbers, such as museums, shopping malls, gyms, churches, and other cultural and public spaces. Yelp, [21] a popular website that provides crowdsourced reviews of local businesses in various countries, added an option whereby its users can identify whether a business has a PokéStop nearby. This allows users to filter the search results of businesses, including restaurants, based on this criterion. We studied the impact of PokéStops on local restaurant reviews that have a PokéStop nearby. We also compared the most recent three years of review trends for these restaurants to analyze the impact of common factors during the period following Pokémon Go's release date.

## 2. RELATED WORK
As the augmented reality industry matures, researchers have studied its socioeconomic impact as well as its impact on users' physical and mental health. Studies to date on Pokémon Go have tried to quantify the socioeconomic effect of the game using survey information. Zach and Tussyadiah [20] analyzed the results from players responding to their survey on likelihood to travel, spend money, and visit restaurants while playing. Overall, the results indicated playing the game encouraged traveling. However, they found players that spent one hour or more playing were less likely to spend money on traveling for the game. Colley et al. [6] combined national field surveys with a geo-statistical analysis of the distribution of PokéStops throughout the United States. Their primary findings indicated that urban areas and areas with large (non-hispanic) white populations were heavily favored in Pokémon Go. Such areas contained far more PokéStops than rural areas and areas with large minority populations. Further, they found that 46% of players had purchased goods at places near where they traveled to play Pokémon Go.

A study done at Microsoft Research by Althoff, White, and Horvitz [1] focused on determining the physical activity involved in playing Pokémon Go by combining signals from large-scale corpora of wearable sensor data and search engine logs for 32,000 users over a period of three months. Pokémon Go players, identified through search engine queries and with activity measured by accelerometry, were found to have a significant increase in physical activity. Based on data collected from the iOS Health app, which records information on the user's physical activity, Howe et al. [9] observed a similar increase in physical activity among both children and adults. Nigg et al. [13] conducted a pre-post study and found among survey recipients that Pokémon Go increased physical activity of various intensity levels by 50 minutes per week and reduced sedentary activity by 30 minutes each day.

Williamson [19] provided anecdotal evidence that suggests numerous users are logging many kilometers in an effort to catch and hatch elusive Pokémon. In addition to referencing claims that Pokémon Go helps its users to fight depression, McCartney [12] reported that the game is playing a positive role in fighting obesity in the U.S. Quinn [16] described how, by remembering the monsters' names and capabilities, children improved their memory capabilities through playing Pokémon GO.

Other studies have looked at Pokémon GO as an opportunity to examine the effect of crowdsourced information on the real world. For example, a study by Wang [18] looked at Pokémon Go as a use case to examine mobile crowd sensing, where communities of users contribute aggregate data generated on their cellphones to make databases of information relevant to the app in a secure fashion. This can be crowdsourced info of where Pokémon were found, such as parking near PokéStops. Wang proposed their own solution for a secure and accurate crowd sensing platform as an answer to the challenges faced by having diversified crowd contributors and sources. Mansilla and Perkis [11] suggested that augmented reality games such as Pokémon Go can contribute to setting up multi-use public spaces called "Adressaparken," that are designed to increase social interaction between community members. Tinati et al. [17] has discussed how the social interaction has been increased by a citizen science game named EyeWire, a game similar to Pokémon Go that challenges its players to map 3D neurons in a retina. Dzodom and Shipman [7] studied how fantasy games play an important role in increasing the social interaction – considered as an important factor by many for maintaining family, friends, and workplace relationships and determining how games play an important role in motivating the players to interact.

Some studies have found Pokémon Go to be a dangerous distraction in some instances. For example, Ayers, Leas, and Dredze [2] processed 4,000 tweets to analyze the distractions attributed to Pokémon Go, with 31–34% of tweets indicating that drivers, passengers, and pedestrians were distracted by the game. Similarly, Joseph and Armstrong [10] showed that players of Pokémon Go tend to be distracted while driving. Pourmand et al. [14] found, by reviewing the PubMed, Medline, and PsycInfo databases, that there was an increase in fractures and dislocations compared to previous video game related injuries. Distracted players injuring themselves was a common theme in the reports.

In a study focused on addressing the safety of players engaged in augmented reality games, Pyae and Potter [15] presented four engagement models designed to capture engagement in the games—"Player," "Play," "Presence," and "Place." The researchers discussed each model based on the individual user scenario and used Pokémon Go as their focal case study. Although there are some reports that Pokémon Go has increased the number of visits made to museums [8] and national parks [3] and also increased sales [4], it is not clear whether the game has had an effect on online reviews of local businesses and restaurants, which is the focus of our analysis.

## 3. DATA

For the present study, all the data pertaining to restaurant reviews were collected from Yelp. Figure 1 shows a Yelp search result that indicates the presence of a PokéStop near a restaurant. Yelp has provided a search filter [5] so that its users can search for restaurants with a PokéStop nearby, as shown in Figure 2. We used the Selenium [22] browser automation framework to navigate the Yelp website and to collect restaurant reviews posted on Yelp. We collected 592,120 reviews from 3,719 distinct local restaurants in 26 U.S. and U.K. cities. We collected the data by running four individual threads: two collected the data for restaurants with a PokéStop nearby, and two for restaurants without a PokéStop in the vicinity.

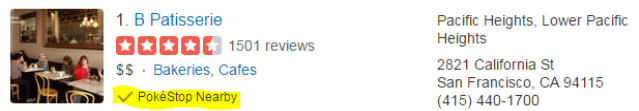

**Figure 1: Yelp search result marking a San Francisco restaurant with a PokéStop nearby.**

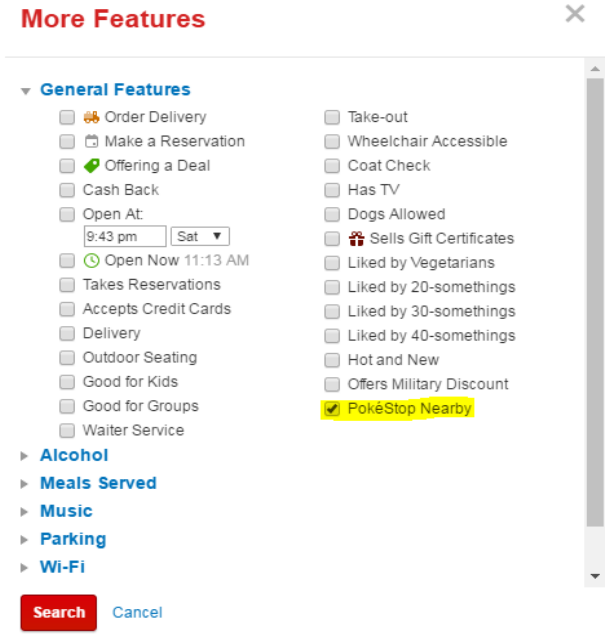

**Figure 2: Yelp's search filter for restaurants and other local businesses with a PokéStop nearby.**

## 4. METHODS

From our two sets of restaurants, those with PokéStops and those without, we narrowed our initial selection by excluding any restaurant that did not have at least five reviews in each month since 2014. From this selection, we took 50 restaurants with the most reviews for each category as our final sample. Thus, we ended with 50 restaurants that were known to have PokéStops nearby and 50 restaurants without, for a total sample of 100 restaurants across both categories. To determine whether the proximity of a PokéStop had any impact on the restaurant reviews, we used the Pearson correlation and Paired T-Test. We split our data points, i.e., the number of reviews year-wise for our two 50 restaurant samples.

We applied the Paired T-Test over another random data samples with 95% confidence. We selected 20 random sample sets (10 samples for restaurants with a PokéStop and 10 samples for those without a PokéStop). Each set consists of 10 randomly selected restaurants from each of the 26 cities, where each city has at least 25 restaurants and each of the restaurants has at least 10 reviews. By applying the previous approach, we eliminated effects of other factors such as weather, special events, etc. We compared the reviews in the months of August to November to see if Pokémon Go affected reviews during that period. August to November were chosen because Yelp added the PokéStop filter in mid-July and our final dataset was collected in the second week of December. As seen in Table 1, we applied tests to compare the previous year, 'Before' period, to the subsequent years 'After' period to measure the change in restaurant reviews.

| Test # | Before | After |
|---|---|---|
| 1 | August-November 2015 | August-November 2016 |
| 2 | August-November 2014 | August-November 2015 |

**Table 1: Data division for the analysis.**

## 5. RESULTS

We studied the impact of Pokémon Go on the restaurants with and without PokéStops for the years 2014 to 2016. We analyzed the reviews from the same period for the previous years, i.e., 2014 and 2015, and found an increase in the number of reviews from 2014 to 2015. To verify these results, we applied the Pearson correlation on data within the stated periods as shown in Table 1. Table 2 shows the results of the Pearson correlation coefficients after performing the correlation analysis. Restaurants with PokéStops nearby had lower correlation compared to that of the restaurants without PokéStops nearby.

| August-November | With PokeStops | Without PokéStops |
|---|---|---|
| 2016 – 2015 | **0.746** | 0.934 |
| 2015 - 2014 | 0.967 | 0.921 |

**Table 2: Pearson correlation coefficients for restaurants with and without PokéStops compared to previous years.**

We then applied the Paired T-Test for test period #1, as indicated in Table 3, which shows that there is no significant difference between the Before and After periods in the case of restaurants with a PokéStop, showing no statistically significant difference in the number of restaurant reviews. However, the test on restaurants without a PokéStop for the same period indicate a significant difference in restaurant reviews for that group. We applied the Paired T-Test for test period #2. The tests show significant differences between the two data sets for the periods of test #2, indicating an increase in the number of reviews from 2014 to 2015.

| Test# | Without PokéStops | With PokéStops |
|-------|-------------------|----------------|
| 1     | 0.0067            | **0.3282**     |
| 2     | 0.0154            | 0.0031         |

**Table 3: P-values for tests #1 and #2.**

## 6. CONCLUSION AND FUTURE WORK

From the previous results, it can be concluded that restaurants with a PokéStop nearby have experienced a slight decrease in the number of reviews on Yelp after the release of Pokémon Go in July 2016, especially in October and November. We intend to study the effect of events delivered by the game's developer, Niantic, Inc., during major holidays to boost user activity in Pokémon Go, on Yelp reviews. We plan to consider the distance between a PokéStop and the nearest restaurant as a factor in the number of reviews posted to Yelp for a given restaurant and whether proximity to a PokéStop has any influence the number of positive and negative reviews on Yelp.